\documentclass[a4paper,11pt]{article}

\usepackage[T2A]{fontenc}
\usepackage[cp1251]{inputenc}
\usepackage[english,ukrainian]{babel}
\usepackage{amssymb,upref}
\usepackage{amsmath,amsthm}
\usepackage{cite}
\usepackage{multicol}
\usepackage{geometry}
\usepackage[dvips]{graphicx}
\usepackage{fancyhdr}
\newtheorem{dfn}{Definition}
\fancypagestyle{plain}{%
\lfoot{\footnotesize \copyright\ С.П.~Майданюк, 2010}%
}

\makeatletter
\renewcommand\section{\@startsection {section}{1}{\z@}%
                                   {-3.5ex \@plus -1ex \@minus -.2ex}%
                                   {2.3ex \@plus.2ex}%
                                   {\normalfont\bfseries}}
\def\@biblabel#1{#1.}
\makeatother

\theoremstyle{plain}

\newtheoremstyle{definition}
  {}
  {}
  {}
  {}
  {\itshape}
  {.}
  {.5em}
  {}
\theoremstyle{definition}

\let\le\leqslant

\begin{document}
\thispagestyle{plain}

УДК 524.85 

\setlength{\columnsep}{7mm}%
\setlength{\columnseprule}{.4pt}
\begin{multicols}{2}

\begin{flushleft}
С.П. Майданюк, к.ф.-м.н., снс\\


{\bfseries Квантовий метод визначення проникності у FRW моделі з радіацією}
\end{flushleft}

{\itshape
У роботі розглядається закрита квантова модель Фрідманна--Робертсона--Уолкера з позитивною космологічною константою, радіацією та газом Чаплигіна. Для аналіза ймовірності народження всесвіту на інфляційній стадії як функції енергії радіації уведено нове визначення хвилі, що ``вільно'' поширюється у сильних полях. Зкорректовано граничну умову тунелювання Віленкіна, обчислено проходження та відбиття у повністю квантовому стаціонарному підході.\\
Ключові слова: фізика раннього всесвіту, інфляція, рівняння Уілера-Де Вітта, газ Чаплигіна.}

\bigskip

{\selectlanguage{english}

\begin{flushleft}
S.P. Maydanyuk, Senior Sc. Researcher\\

{\bfseries Quantum method of determination of penetrability in FRW model with radiation}
\end{flushleft}

{\itshape
In paper the closed Friedmann--Robertson--Walker model with quantization in presence of the positive cosmological constant, radiation and Chaplygin gas is studied. For analysis of tunneling probability for birth of an asymptotically deSitter, inflationary Universe as a function of the radiation energy a new definition of a ``free'' wave propagating inside strong fields is introduced. Vilenkin's tunneling boundary condition is corrected, penetrability and reflection are calculated in fully quantum stationary approach.

\vfill
}}

\end{multicols}

E-mail: maidan@kinr.kiev.ua

\setlength{\columnsep}{4mm}%
\setlength{\columnseprule}{0pt}
\begin{multicols}{2}
\section{Introduction}
\vspace{-3mm}

To date, among all variety of models of early Universe one can select two prevailing approaches: these are the Feynman formalism of path integrals in multidimensional spacetime, developed by the Cambridge group and other researchers, called the \emph{``Hartle--Hawking method''}~\cite{Hartle.1983.PRD}, and a method based on direct consideration of tunneling in 4-dimensional Euclidian spacetime called the \emph{``Vilenkin method''}~\cite{Vilenkin.1982.PLB}. In the quantum approach we have the following picture of the Universe creation: a closed Universe with a small size is formed from ``nothing'' (vacuum), where by the word ``nothing'' one refers to a quantum state without classical space and time. A wave function is used for a probabilistic description of the creation of the Universe and such a process is connected with transition of a wave through an effective barrier.

In majority of models tunneling is studied in details in the semiclassical approximations (for example, see \cite{Vilenkin.1994.PRD,Rubakov.1999,Casadio.2005.PRD.D71,Luzzi.2006.PhD}). Here, a \emph{tunneling boundary condition} \cite{Vilenkin.1994.PRD} could seems to be natural, where the wave function should represent an outgoing wave at the large scale factor $a$. However, whether is such a wave free in asymptotic region? If to draw attention on increasing gradient of potential, used with opposite sign and having a sense of force, acting ``through the barrier'' on this wave, then one come to contradiction: \emph{influence of the potential on this wave is increased strongly at increasing $a$} \cite{Maydanyuk.2010.IJMPD}. Now a new question has been appeared: what should the wave represent in the cosmological problem? So, we come to necessity \emph{to define ``free'' wave inside strong fields}.

The problem of correct definition of the wave in cosmology is reinforced more, if to calculate incident and reflected waves before the barrier. Even with known exact solution for the wave function there is uncertainty in determination of these waves. But penetrability is based on them.

In order to estimate probability of formation of Universe accurately, we need in the fully quantum basis.
%
Aims of this paper are:
(1) to give definition of the wave in strong fields;
(2) to construct the fully quantum stationary method of determination of the penetrability and reflection using the definition of the wave above;
(3) to estimate how much the semiclassical approach is differed from the fully quantum one.

\section{A model in the Friedmann--Robertson--Walker metric with radiation and generalized Chaplygin gas
\label{sec.model}}

Let us start from a case of a closed ($k=1$) FRW model in the presence of a positive cosmological constant $\Lambda > 0$, radiation and component of the Chaplygin gas. The minisuperspace Lagrangian has the form~\cite{Maydanyuk.2008.EPJC}:
\begin{equation}
\begin{array}{lcl}
  \mathcal{L}\,(a,\dot{a}) =
  \displaystyle\frac{3\,a}{8\pi\,G}\:
  \biggl(-\dot{a}^{2} + k - \displaystyle\frac{8\pi\,G}{3}\; a^{2}\,\rho(a) \biggr),
\end{array}
\label{eq.model.3.8}
\end{equation}
where
$a$ is scale factor,
$\dot{a}$ is derivative of $a$ with respect to time coordinate $t$,
$\rho\,(a)$ is a generalized energy density.
In order to connect the stage of Universe with dust matter and its another accelerating stage, in Ref.~\cite{Kamenshchik.2001.PLB} a new scenario with the \emph{Chaplygin gas} was applied to cosmology. A quantum FRW-model with the Chaplygin gas has been constructed on the basis of equation of state instead of $p\,(a)=\rho\,(a)/3$ (where $p\,(a)$ is pressure)
by
$ p_{\rm Ch} = - A / \rho_{\rm Ch}^{\alpha}$,
where $A$ is positive constant and $0< \alpha \le 1$.
Solution of equation of state gives
\begin{equation}
  \rho_{\rm Ch}(a) = \biggl( A + \displaystyle\frac{B}{a^{3\,(1+\alpha)}} \biggr)^{1/(1+\alpha)},
\label{eq.7.1.2}
\end{equation}
where $B$ is a new constant. This model through one phase $\alpha$ connects the stage of Universe where dust dominates and DeSitter stage. At limit $\alpha \to 0$ eq.~(\ref{eq.7.1.2}) is transformed into $\rho_{\rm dust}$ plus $\rho_{\Lambda}$. From such limit we find
$A = \rho_{\Lambda}$, $B = \rho_{\rm dust}$
and write the following generalized density \cite{Maydanyuk.2010.IJMPD}:
\begin{equation}
\begin{array}{lcl}
  \rho\,(a) =
    \biggl( \rho_{\Lambda} + \displaystyle\frac{\rho_{\rm dust}}{a^{3\,(1+\alpha)}} \biggr)^{1/(1+\alpha)} +
    \displaystyle\frac{\rho_{\rm rad}}{a^{4}(t)}, &
\end{array}
\label{eq.7.1.4}
\end{equation}
where $\rho_{\rm rad}(a)$ is component describing the radiation (equation of state for radiation is $p(a)=\rho_{\rm rad}(a)/3$, $p$ is pressure) and $\rho_{\Lambda} = \Lambda / (8\pi\,G)$.

The passage to the quantum description of the evolution of the Universe is obtained by the procedure of canonical quantization in the Dirac formalism for systems with constraints. We obtain the Wheeler--De Witt (WDW) equation, which after multiplication on factor and passage of component with radiation $\rho_{\rm rad}$ into right part transforms into the following form \cite{Maydanyuk.2010.IJMPD}:
\begin{equation}
\begin{array}{cc}
  \biggl\{ -\:\displaystyle\frac{\partial^{2}}{\partial a^{2}} + V\,(a) \biggr\}\; \varphi(a) =
  E_{\rm rad}\; \varphi(a), &
  E_{\rm rad} = \displaystyle\frac{3\,\rho_{\rm rad}}{2\pi\,G},
\end{array}
\label{eq.7.2.1}
\end{equation}
\begin{equation}
\begin{array}{ccl}
  V\,(a) & = &
    \biggl( \displaystyle\frac{3}{4\pi\,G} \biggr)^{2}\: k\,a^{2} -
    \displaystyle\frac{3}{2\pi\,G}\:
     a^{4}\, \times \\
  & \times &
    \biggl( \rho_{\Lambda} + \displaystyle\frac{\rho_{\rm dust}}{a^{3\,(1+\alpha)}} \biggr)^{1/(1+\alpha)},
\end{array}
\label{eq.7.2.2}
\end{equation}
where $\varphi(a)$ is wave function of Universe.
For the Universe of closed type (at $k=1$) at $8\pi\,G \equiv M_{\rm p}^{-2} = 1$ we have:
\begin{equation}
\begin{array}{cc}
  V\,(a) =
    36\,a^{2} -
    12\,a^{4}\,\Bigl(\Lambda + \displaystyle\frac{\rho_{\rm dust}}{a^{3\,(1+\alpha)}} \Bigr)^{1/(1+\alpha)} 
\end{array}
\label{eq.7.2.3}
\end{equation}
and $E_{\rm rad} = 12\, \rho_{\rm rad}$.

Let us expand the potential (\ref{eq.7.2.3}) close to arbitrary point $\bar{a}$ by powers $q=a-\bar{a}$
and restrict ourselves by linear item:
\begin{equation}
  V_{\rm Ch}\,(q) = V_{0} + V_{1}q.
\label{eq.7.3.1}
\end{equation}
For coefficients $V_{0}$ and $V_{1}$ I find:
\begin{equation}
\begin{array}{ccl}
  V_{0} & = & V_{\rm Ch}\,(a=\bar{a}), \\
  V_{1} & = &
    72\,a +
    12\,a^{3}\,
    \Bigl\{ -4\,\Lambda - \displaystyle\frac{\rho_{\rm dust}}{a^{3\,(1+\alpha)}} \Bigr\} \times \\
  & \times &
    \Bigl( \Lambda + \displaystyle\frac{\rho_{\rm dust}}{a^{3\,(1+\alpha)}} \Bigr)^{-\alpha/(1+\alpha)}
\end{array}
\label{eq.7.3.3}
\end{equation}
and eq.~(\ref{eq.7.2.1}) obtains form:
\begin{equation}
  -\displaystyle\frac{d^{2}}{dq^{2}}\, \varphi(q) + (V_{0} - E_{\rm rad} + V_{1}\, q)\: \varphi(q) = 0.
\label{eq.7.3.4}
\end{equation}
After change of variable
\begin{equation}
  \xi = \displaystyle\frac{E_{\rm rad} - V_{0}} {|V_{1}|^{2/3}} - \displaystyle\frac{V_{1}} {|V_{1}|^{2/3}}\: q
\label{eq.7.3.10}
\end{equation}
we have
\begin{equation}
  \displaystyle\frac{d^{2}}{d\xi^{2}}\, \varphi(\xi) + \xi\, \varphi(\xi) = 0.
\label{eq.7.3.9}
\end{equation}

\section{Motivations to correct Vilenkin's boundary condition of tunneling
\label{sec.3}}

Let us analyze how much a choice of the boundary condition in the asymptotic region is motivated.
\begin{itemize}
\item
In tasks of decay in nuclear and atomic physics the potentials of interactions tend to zero in the asymptotic region. Here, an application of the boundary condition at limit of infinity does not give questions. In cosmology we deal with another, principally different type of the potential: with increasing of the scale factor $a$ modulus of this potential increases. A gradient of the potential, used with opposite sign and having a sense of force acting on the wave, increases also. So, there is nothing mutual with free propagation of the wave in the asymptotic region \cite{Maydanyuk.2010.IJMPD}.

\item
Results of Ref.~\cite{Maydanyuk.2008.EPJC} reinforce a seriousness of such a problem: the scale factor $a$ in the external region is larger, the period of oscillations of the wave function is smaller. This requires with increasing $a$ to decrease step of calculations of the wave function. This increases time of calculations, increases errors. So, boundary condition in the asymptotic region has no practical sense in cosmology. In contrary, in nuclear and atomic physics calculations of the asymptotic wave are the most stable.

\item
It has not been known whether Universe expands at extremely large scale factor $a$. Just the contrary, it would like to clarify this, imposing Universe to expand in initial stage.
\end{itemize}
So, I redefine the boundary condition as \cite{Maydanyuk.2010.IJMPD}
\begin{quote}
\emph{the boundary condition should fix the wave function so that it represents the wave, interaction between which and the potential barrier is minimal at such a value of the scale factor $a$ where action of this potential is minimal.}
\end{quote}
To give a mathematical formulation for this definition, we are confronted with two questions:

\begin{enumerate}
\item
What should the free wave represent at arbitrary point inside cosmological potential?

\item
At which coordinate is imposition of this boundary condition the most corrected?
\end{enumerate}

At first, let us solve the second question. Which should this point be: or this is a turning point (where the potential coincides with energy of radiation), or this is a coordinate where a gradient of the potential (having a sense of \emph{force of interaction}) becomes zero, or this is a coordinate where the potential becomes zero?
\begin{quote}
\emph{We define this coordinate where the force acting on the wave is minimal. We define the force as the gradient of the potential used with opposite sign.}
\end{quote}

\section{Definition of the wave minimally interacting with the potential
\label{sec.4}}

\begin{dfn}[strict definition of the wave]
\label{def.wave.strict}
The wave is such a linear combination of two partial solutions of the wave function that the change of the modulus $\rho$ of this wave function is closest to constant under variation of $a$:
\begin{equation}
  \displaystyle\frac{d^{2}}{da^{2}}\, \rho(a) \biggl|_{a=a_{tp}} \to 0.
\label{eq.5.2.4}
\end{equation}
\end{dfn}
\noindent
For some types of potentials it is more convenient to define the wave less strongly.

\begin{dfn}[weak definition of wave]
The wave is such a linear combination of two partial solutions of wave function that the modulus $\rho$ changes minimally under variation of $a$:
\begin{equation}
  \displaystyle\frac{d}{da}\, \rho(a) \biggl|_{a=a_{tp}} \to 0.
\label{eq.5.2.5}
\end{equation}
\end{dfn}

Now we shall look for the function $\varphi(\xi)$ as
\begin{equation}
  \varphi\, (\xi) = T \cdot \Psi^{(+)}(\xi),
\label{eq.5.2.6}
\end{equation}
\begin{equation}
\begin{array}{ccl}
  \Psi^{(\pm)} (\xi) & = &
    \displaystyle\int\limits_{0}^{u_{\rm max}}
    \exp{\pm\,i\,\Bigl(-\displaystyle\frac{u^{3}}{3} + f(\xi)\,u \Bigr)} \; du,
\end{array}
\label{eq.5.2.7}
\end{equation}
where $T$ is an unknown factor, $f(\xi)$ is an unknown continuous function satisfying $f(\xi) \to {\rm const}$ at $\xi \to 0$, and $u_{\rm max}$ is the unknown upper limit of integration. The real part of $f(\xi)$ gives a contribution to the phase of the integrand function while the imaginary part of $f(\xi)$ deforms modulus.

At small enough values of $|\xi|$ we represent $f(\xi)$ in the form of a power series:
\begin{equation}
  f(\xi) = \sum\limits_{n=0}^{+\infty} f_{n}\, \xi^{n},
\label{eq.5.2.10}
\end{equation}
where $f_{n}$ are constants. Substituting formula (\ref{eq.5.2.7}) for $\Psi(\xi)$ into equation (\ref{eq.7.3.9}), we find (see Ref.~\cite{Maydanyuk.2010.IJMPD}):
\begin{equation}
\begin{array}{lcl}
  f_{2}^{(\pm)} = \pm\;\displaystyle\frac{f_{1}^{2}}{2i} \cdot \displaystyle\frac{J_{2}^{(\pm)}}{J_{1}^{(\pm)}}, \\
  f_{3}^{(\pm)} =
    \pm\;\displaystyle\frac{4 f_{1}f_{2}^{(\pm)}\, J_{2}^{(\pm)} - J_{0}^{(\pm)}} {6i\, J_{1}^{(\pm)}}, \\
  f_{n+2}^{(\pm)} =
    \displaystyle\frac{\sum\limits_{m=0}^{n} (n-m+1)(m+1) \: f_{n-m+1}^{(\pm)}\,f_{m+1}^{(\pm)}}
      {i \: (n+1)(n+2)} \cdot
    \displaystyle\frac{J_{2}^{(\pm)}} {J_{1}^{(\pm)}},
\end{array}
\label{eq.5.2.16}
\end{equation}
where
\begin{equation}
\begin{array}{lcl}
  J_{0}^{(\pm)} =
    \displaystyle\int\limits_{0}^{u_{\rm max}}
    \exp{\pm i\, \Bigl(-\displaystyle\frac{u^{3}}{3} + f_{0}u \Bigr)} \; du, \\
  J_{1}^{(\pm)} =
    \displaystyle\int\limits_{0}^{u_{\rm max}}
    u \: \exp{\pm i\, \Bigl(-\displaystyle\frac{u^{3}}{3} + f_{0}u \Bigr)} \; du, \\
  J_{2}^{(\pm)} =
    \displaystyle\int\limits_{0}^{u_{\rm max}}
    u^{2} \: e^{\pm i\, \Bigl(-\displaystyle\frac{u^{3}}{3} + f_{0}u \Bigr)} \; du.
\end{array}
\label{eq.5.2.17}
\end{equation}
In order to be the solution $\Psi(\xi)$ closer to the well-known Airy functions, ${\rm Ai}\,(\xi)$ and ${\rm Bi}\,(\xi)$, we choose $f_{0} = 0$, $f_{1} = 1$.

\section{Calculations of the wave function
\label{sec.5}}

In order to provide a linear independence between two partial solutions for the wave function effectively, I look for the first partial solution increasing in the region of tunneling and the second one decreasing in this tunneling region. At first, I define each partial solution and its derivative at a selected starting point, and then I calculate them in the region close enough to this point using the \emph{method of beginning of the solution}. Here, for the partial solution which increases in the barrier region, as the starting point I use arbitrary point $\bar{a}$ inside well with its possible shift at non-zero energy $E_{\rm rad}$ or equals to zero $a=0$ at null energy $E_{\rm rad}$, and for the second partial solution which decreases in the barrier region, I select the starting point to be equal to external turning point $a_{\rm tp,\, out}$. Then both partial solutions and their derivatives I calculate independently in the whole required range of $a$ using the \emph{method of continuation of the solution}, which is improvement of the Numerov method with a constant step. By such a way, I obtain two partial solutions for the wave function and their derivatives in the whole studied region.

Having obtained two linearly independent partial solutions $\varphi_{1}(a)$ and $\varphi_{2}(a)$,
we make up a general solution (prime is for derivative with respect to $a$):
\begin{equation}
  \varphi\,(a) = T \cdot \bigl(C_{1}\, \varphi_{1}(a) + C_{2}\,\varphi_{2}(a) \bigr),
\label{eq.5.3.1}
\end{equation}
\begin{equation}
\begin{array}{cc}
  C_{1} = \displaystyle\frac{\Psi\varphi_{2}^{\prime} - \Psi^{\prime}\varphi_{2}}
          {\varphi_{1}\varphi_{2}^{\prime} - \varphi_{1}^{\prime}\varphi_{2}} \bigg|_{\bar{a}}, &
  C_{2} = \displaystyle\frac{\Psi^{\prime}\varphi_{1} - \Psi\varphi_{1}^{\prime}}
          {\varphi_{1}\varphi_{2}^{\prime} - \varphi_{1}^{\prime}\varphi_{2}} \bigg|_{\bar{a}},
\end{array}
\label{eq.5.3.3}
\end{equation}
where $T$ is normalization factor, $C_{1}$ and $C_{2}$ are constants found from the boundary condition.

\section{Problem of interference between the incident and reflected waves
\label{sec.6}}

Rewriting the wave function $\varphi_{\rm total}$ in the internal region through a summation of incident $\varphi_{\rm inc}$ wave
and reflected $\varphi_{\rm ref}$ wave:
\begin{equation}
  \varphi_{\rm total} =
  \varphi_{\rm inc} + \varphi_{\rm ref},
\label{eq.4.1.1}
\end{equation}
we consider the total flux:
\begin{equation}
\begin{array}{ccc}
  j\, (\varphi_{\rm total}) & = &
  j_{\rm inc} + j_{\rm ref} + j_{\rm mixed},
\end{array}
\label{eq.4.1.2}
\end{equation}
where
\begin{equation}
\begin{array}{lcl}
  j_{\rm inc} =
    i\, \Bigl(\varphi_{\rm inc} \nabla \varphi_{\rm inc}^{*} - \mbox{h.~c.}\Bigr), \\
  j_{\rm ref} =
    i\, \Bigl(\varphi_{\rm ref} \nabla \varphi_{\rm ref}^{*} - \mbox{h.~c.}\Bigr), \\
  j_{\rm mixed} =
    i\, \Bigl(
      \varphi_{\rm inc} \nabla \varphi_{\rm ref}^{*} +
      \varphi_{\rm ref} \nabla \varphi_{\rm inc}^{*} - \mbox{h.~c.}
    \Bigr).
\end{array}
\label{eq.4.1.3}
\end{equation}
The $j_{\rm mixed}$ component describes interference between the incident and reflected waves in the internal region (let us call it as \emph{mixed component of the total flux} or simply \emph{flux of mixing}).
From constancy of the total flux $j_{\rm total}$ we find flux $j_{\rm tr}$ for the wave
transmitted through the barrier, and:
\begin{equation}
\begin{array}{cc}
  j_{\rm inc} = j_{\rm tr} - j_{\rm ref} - j_{\rm mixed}, &
  j_{\rm tr} = j_{\rm total} = {\rm const}.
\end{array}
\label{eq.4.1.5}
\end{equation}
Now one can see that \emph{the mixed flux introduces ambiguity in determination of the penetrability and reflection for the same known wave function.}

In the radial problem of quantum decay definition of penetrability and reflection looks to be conditional as the incident and reflected waves should be defined inside internal region from the left of the barrier. In order to formulate these coefficients, we shall include into definitions coordinates where the fluxes
are defined (denote them as $x_{\rm left}$ and $x_{\rm right}$):
\begin{equation}
\begin{array}{ccc}
\hspace{-2mm}
  T = \displaystyle\frac{j_{\rm tr}(x_{\rm right})}{j_{\rm inc}(x_{\rm left})}, &
\hspace{-2mm}
  R = \displaystyle\frac{j_{\rm ref}(x_{\rm left})}{j_{\rm inc}(x_{\rm left})}, &
\hspace{-2mm}
  M = \displaystyle\frac{j_{\rm mixed}(x_{\rm left})}{j_{\rm inc}(x_{\rm left})}.
\end{array}
\label{eq.4.2.1}
\end{equation}
From eqs.~(\ref{eq.4.1.5}) and (\ref{eq.4.2.1}) we obtain \cite{Maydanyuk.2010.IJMPD}
($j_{\rm tr}$ and $j_{\rm ref}$ are directed in opposite directions,
$j_{\rm inc}$ and $j_{\rm tr}$  --- in the same directions):
\begin{equation}
  |T| + |R| - M = 1.
\label{eq.4.2.3}
\end{equation}
\emph{Now we see that condition
$|T| + |R| = 1$
has sense in quantum mechanics only if there are no any interference between incident and reflected waves which we calculate}, and it is to use
$j_{\rm mixed} = 0$.

\section{The penetrability and reflection: the fully quantum approach versus semiclassical one
\label{sec.7}}

Now we shall estimate by the method described above the coefficients of penetrability and reflection for the potential barrier (\ref{eq.7.2.3}) with parameters $A=36$, $B=12\,\Lambda$, $\Lambda=0.01$ at different values of the energy of radiation $E_{\rm rad}$. We shall compare the found coefficient of penetrability with its value, which the semiclassical method gives.
In the semiclassical approach we shall consider two following definitions of this coefficient:
\begin{equation}
\begin{array}{lcl}
\vspace{4mm}
  P_{\rm penetrability}^{\rm WKB, (1)} & = & \displaystyle\frac{1}{\theta^{2}}, \\
  P_{\rm penetrability}^{\rm WKB, (2)} & = & \displaystyle\frac{4}{\Bigl(2\theta + 1/(2\theta)^{2}\Bigr)^{2}},
\end{array}
\label{eq.6.1}
\end{equation}
where
\begin{equation}
  \theta =
    \exp \displaystyle\int\limits_{a_{\rm tp}^{\rm (int)}}^{a_{\rm tp}^{\rm (ext)}} \bigl|V(a)-E\bigr|\; da.
\label{eq.6.2}
\end{equation}
One can estimate also \emph{duration of a formation of the Universe},
using by definition (15) in Ref.~\cite{Monerat.2007.PRD}:
\begin{equation}
  \tau = 2\, a_{\rm tp,\, int}\: \displaystyle\frac{1}{\rm P_{penetrability}}.
\label{eq.6.3}
\end{equation}

\end{multicols}
\begin{table}
\hspace{-20mm}
\begin{center}
\begin{tabular}{|c|c|c|c|c|c|} \hline
  Energy
  & \multicolumn{3}{|c|}{Penetrability $P_{\rm penetrability}$}
  & \multicolumn{2}{|c|}{Time $\tau$} \\ \cline{2-6}
  & Full QM method 1 & Full QM method 2 & Method WKB & Full QM method 1 & Method WKB \\ \hline
 20.0 &  $7.6221 \times 10^{-30}$ & $7.7349 \times 10^{-29}$ & $1.4692 \times 10^{-29}$ &  $9.4313 \times 10^{+29}$ &  $4.8928 \times 10^{+29}$ \\
 40.0 &  $3.5680 \times 10^{-26}$ & $6.5169 \times 10^{-25}$ & $1.2298 \times 10^{-25}$ &  $2.1257 \times 10^{+26}$ &  $6.1670 \times 10^{+25}$ \\
 60.0 &  $2.0591 \times 10^{-22}$ & $4.3423 \times 10^{-21}$ & $8.1523 \times 10^{-22}$ &  $3.8719 \times 10^{+22}$ &  $9.7797 \times 10^{+21}$ \\
 80.0 &  $1.5530 \times 10^{-18}$ & $2.3850 \times 10^{-17}$ & $4.4346 \times 10^{-18}$ &  $5.3862 \times 10^{+18}$ &  $1.8862 \times 10^{+18}$ \\
100.0 &  $3.2922 \times 10^{-14}$ & $1.1053 \times 10^{-13}$ & $2.0304 \times 10^{-14}$ &  $2.6622 \times 10^{+14}$ &  $4.3167 \times 10^{+14}$ \\
120.0 &  $8.6052 \times 10^{-11}$ & $4.4005 \times 10^{-10}$ & $7.9523 \times 10^{-11}$ &  $1.0678 \times 10^{+11}$ &  $1.1555 \times 10^{+11}$ \\
140.0 &  $2.2012 \times 10^{-8}$ & $1.5460 \times 10^{-6}$ & $2.7128 \times 10^{-7}$ &  $4.3888 \times 10^{+8}$ &  $3.5612 \times 10^{+7}$ \\
160.0 &  $2.9685 \times 10^{-5}$ & $4.9980 \times 10^{-3}$ & $8.1663 \times 10^{-4}$ &  $3.4471 \times 10^{+5}$ &  $1.2530 \times 10^{+4}$ \\
170.0 &  $3.4894 \times 10^{-3}$ & $2.6078 \times 10^{-1}$ & $4.2919 \times 10^{-2}$ &  $3.0460 \times 10^{+3}$ &  $2.4820 \times 10^{+2}$ \\
\hline
\end{tabular}
\end{center}
\caption{\small
The penetrability $P_{\rm penetrability}$ of the barrier and duration $\tau$ of the formation of the Universe defined by eq.~(\ref{eq.6.3}) in the FRW-model with the Chaplygin gas obtained in the fully quantum and semiclassical approaches
(minimum of the hole is -93.579 and its coordinate is 1.6262,
maximum of the barrier is 177.99 and its coordinate is 5.6866):
the fully QM method 1 is calculations by the fully quantum approach for the boundary located in the coordinate of the minimum of the internal hole (i.~e. coordinate is 1.6262),
the fully QM method 2 is calculations by the fully quantum approach for the boundary located in the internal turning point $a_{\rm tp,\, in}$ (coordinates of the turning points are in Tabl.~2)
\label{table.1}}
\end{table}
\begin{multicols}{2}

Results are presented in Tabl.~\ref{table.1}. $P_{\rm penetrability}^{\rm WKB, (2)}$ is not included in the table because it coincidence with $P_{\rm penetrability}^{\rm WKB, (1)}$ to the first 8 digits. One can see that the fully quantum approach gives the penetrability close to its semiclassical value, which differs from results~\cite{Monerat.2007.PRD}.

In the next Tabl.~2 the coefficients of the penetrability, reflection and mixing calculated in the fully quantum method are presented for the energy of radiation $E_{\rm rad}$ close to the height of the barrier. One can see that summation of all such values for coefficients allows to reconstruct the property (\ref{eq.4.2.3})
with accuracy of the first 11--18 digits.
Now it becomes clear that the approach proposed in Ref.~\cite{Monerat.2007.PRD} and the semiclassical methods do not give such an accuracy.

\end{multicols}
\begin{table}
\hspace{-20mm}
\begin{center}
\begin{tabular}{|c|c|c|c|c|c|c|c|} \hline
  Energy & \multicolumn{4}{|c|}{Fully quantum method} &
  \multicolumn{2}{|c|}{Turning points} \\ \cline{2-7} &
  Penetrability &
  Reflection &
  Interference &
  Summation &
  $a_{\rm tp,\, in}$ & $a_{\rm tp,\, out}$ \\ \hline
 20.0 &  $7.6221543404 \times 10^{-30}$ &  1.00000000000000 &  $8.55 \times 10^{-20}$ &  1.00000000000000 &  3.59 &  7.05 \\
 40.0 &  $3.5680158760 \times 10^{-26}$ &  1.00000000000000 &  $2.34 \times 10^{-19}$ &  1.00000000000000 &  3.79 &  6.97 \\
 60.0 &  $2.0591415452 \times 10^{-22}$ &  1.00000000000000 &  $4.86 \times 10^{-20}$ &  1.00000000000000 &  3.98 &  6.88 \\
 80.0 &  $1.5530040238 \times 10^{-18}$ &  1.00000000000000 &  $2.08 \times 10^{-19}$ &  1.00000000000000 &  4.18 &  6.78 \\
100.0 &  $3.2922846164 \times 10^{-14}$ &  0.99999999999996 &  $3.13 \times 10^{-20}$ &  1.00000000000000 &  4.38 &  6.67 \\
120.0 &  $8.6052092530 \times 10^{-11}$ &  0.99999999991394 &  $1.06 \times 10^{-19}$ &  1.00000000000000 &  4.59 &  6.55 \\
140.0 &  $2.2012645564 \times 10^{-8}$ &  0.99999997798735 &  $1.60 \times 10^{-19}$ &  1.00000000000000 &  4.83 &  6.39 \\
160.0 &  $2.9685643504 \times 10^{-5}$ &  0.99997031435611 &  $6.94 \times 10^{-20}$ &  0.99999999999961 &  5.11 &  6.18 \\
170.0 &  $3.4894544195 \times 10^{-3}$ &  0.99651054553176 &  $2.02 \times 10^{-19}$ &  0.99999999995131 &  5.31 &  6.02 \\
\hline
\end{tabular}
\end{center}
\caption{\small The coefficients of the penetrability, reflection and mixing calculated by the fully quantum method and test on their summation for the FRW-model with the Chaplygin gas density component (the fully quantum approach 1 is used at the internal boundary located in the coordinate of the minimum of the internal hole)
\label{table.4}}
\end{table}
\begin{multicols}{2}


\section{Conclusions
\label{sec.conclusions}}

In the paper the closed Friedmann--Robertson--Walker model with quantization in the presence of a positive cosmological constant, radiation and Chaplygin gas is studied. Note the following.
\begin{enumerate}
\item
A fully quantum definition of the wave propagating inside strong field and interacting with them minimally has been formulated. The tunneling boundary condition has been corrected.

\item
A quantum stationary method of determination of penetrability and reflection relatively the barrier has been developed. Here, non-zero interference between the incident and reflected waves has been taken into account and for its estimation the coefficient of mixing has been introduced.
\end{enumerate}
In such an approach the penetrability of the barrier for the studied FRW-model has been estimated. Note the following.
\begin{itemize}
\item
The probability for birth of asymptotically deSitter Universe is close to results obtained by the semiclassical approach, but differs on results obtained by non-stationary approach~\cite{Monerat.2007.PRD} (see Tabl.~1 and 2).

\item
The reflection from the barrier has been determined at first time. It is differed essentially on 1 at the energy of radiation close enough to the barrier height (see Tabl.~2).

\item
The modulus of the coefficient of mixing (indicating interference between the incident and reflected waves) is less $10^{-19}$.

\item
A property (\ref{eq.4.2.3}) is reconstructed to the first 11--18 digits (see Tabl.~2).
\end{itemize}
%


\renewcommand{\refname}{References}


\begin{flushright}
Надійшла до редколегії 0X.0X.2010
\end{flushright}
\end{multicols}

\end{document}